# A Quick Look at Serial ATA (SATA) Disk Performance

Tom Barclay

Wyman Chong

Jim Gray

October 2003

Microsoft Research

Microsoft Corporation

One Microsoft Way

Redmond, WA  98052

# A Quick Look at SATA Disk Performance

Tom Barclay, Wyman Chong, Jim Gray
{TBarclay, WymanC, Gray }@microsoft.com

Microsoft Research, 455 Market St., Suite 1690, San Francisco, CA 94105
http://research.microsoft.com/barc

*Abstract*

We have been investigating the use of low-cost, commodity components for multi-terabyte SQL Server databases [SQL]. Dubbed *storage bricks*, these servers are white box PCs containing the largest ATA drives, value-priced AMD or Intel processors, and inexpensive ECC memory. One issue has been the wiring mess, air flow problems, length restrictions, and connector failures created by seven or more parallel ATA (PATA) ribbon cables and drives in]a tower or 3U rack-mount chassis. Large capacity Serial ATA (SATA) drives have recently become widely available for the PC environment at a reasonable price. In addition to being faster, the SATA connectors seem more reliable, have a more reasonable length restriction (1m) and allow better airflow. We tested two drive brands along with two RAID controllers to evaluate SATA drive performance and reliablility. This paper documents our results so far.

## 1 Overview

We built and operate the TerraServer-USA web site – http://terraserver-usa.com [Barclay98, Barclay99, Barclay00, Barclay02]. It has been on the Internet since June 1998, and is one of the world's largest public repositories of high-resolution aerial and topographic data. The TerraServer hosts about 3.3 terabytes of user data – growing about 20GB per month as new data arrives from the USGS [NATLMAP03]. The imagery is compressed into Jpeg or GIF format and stored in SQL Server relational databases [SQL].

Since September 2000, the TerraServer-USA hardware environment has been a classic three tier application (DMZ, web server, database server) backed by an enterprise class Storage Area Network (SAN.) 18 terabytes of raw disk storage is connected to the four database servers using fiber-channel. All the servers and storage were supplied by Hewlett-Packard.

The system is designed for high availability. The database servers use Microsoft Clustering Service (MSC) with a hot standby server [MSCS03]. The storage environment has triple-mirrored disks, dual redundant controllers and dual-redundant SAN switch fabrics. This configuration has performed very well and has had no service outages – other than server failovers that stall part of the database for less than a minute. When built in September 2000, this configuration including compute servers had a list price of $1.6 million.

We are building the next generation TerraServer-USA environment using storage bricks. The goal of the project is to meet or exceed the performance and availability achieved by the SAN environment over the last three years at 10% of the capital cost, and a 1/3 the operations cost (hosting charges and operations tasks).

ATA drives typically cost 800$/TB (PATA or Parallel ATA) and 1k$/TB (SATA or Serial ATA). Controllers, power, cooling, processors, and sheet metal bring the "packaged" price up to about 2k$/TB (or 3k$/TB for branded systems).

We experimented with a number of hardware configurations and database partitioning schemes. A seven disk 1.75 TB storage brick – with one 2.4 GHz processor and 1 GB of RAM can comfortably support the database activity of 1.4 TB of TerraServer data. So, 3 such storage bricks costing about $10,000 could support the TerraServer backend database activity. But, such a design would have no fault-tolerance.

The experiments reported in this document verify that two hyper-threaded 2.4 GHz processors with 4 GB of RAM, and 16 disks (2 TB mirrored) can comfortably support the load and tolerate disk failures.

We plan to deploy 3 such machines to support the TerraServer load, with an addition 3 mirrors and a spare server (7 nodes in all). This design is highly fault tolerant, and has a very easy recovery and management model. A separate paper will describe that design once the systems are operational.

We commissioned Silicon Mechanics [SM03], a white-box PC manufacturer in Seattle Washington, to build two Storage Bricks with the following common attributes:
- 3u rack mountable chassis
- Dual 2.4ghz Xeon processors with hyper-threading
- SuperMicro Motherboard
- 4 GB of RAM
- 3ware 8506 SATA disk controllers
- 8 Western Digital 250gb SATA drives each with 8mb of cache
- 8 Maxtor 250gb SATA drives with 8mb of cache
- 2 on-board 10/100/1000 ethernet NIC
- 5 amps @ 208 volts

One Storage Brick was configured with two 8 port SATA RAID controllers manufactured by 3Ware [3WARE]. The second Storage Brick was configured with four 4 port SATA RAID controllers manufactured by 3Ware. At the time the systems were configured, there were no other SATA RAID controllers available that support the 64-bit PCI bus.

The experiment was architected to answer four high level questions:
1. Which SATA drive performs best or is more reliable?
2. What are the performance differences between two 4-port controllers and an 8-port controller?
3. What are the performance and manageability differences between hardware and software RAID1 (mirroring)?
4. What percentage of the normal peak weekly TerraServer application load can a storage brick support?

## 2 Test Environment

### 2.1 Test Methods

We used two methods to evaluate the performance of SATA disks and controllers – stress testing a single volume and then running the storage bricks under the simulated load of the real Teraserver.

#### 2.1.1 Stress Test a Single Volume

The first method was designed to measure the maximum throughput of the drives and the controllers using synthetic I/O profiles. We used the SQLIO.EXE program originally written by the lead SQL Server developer to asses the performance of a disk sub-system running SQL Server.

SQLIO can be configured to and measure read versus write, and sequential versus random performance varying the number of memory buffers, number overlapped (parallel) I/O operations, etc. We configured the SQLIO tests on one volume as follows:
- Sequential read using 64Kb buffers and 4-deep I/O.
- Sequential write using 64Kb buffers and 4-deep I/Os.
- Random read using 8Kb buffers and 1-deep I/O (no overlapped I/O)
- Random write using 8Kb buffers and 1-dep I/O.
- Random read using 8Kb buffers and 4-deep I/Os.
- Random write using 8Kb buffers and 4-deep I/Os.

We found that there was no perceptible difference in performance when running a test for 1 minute versus 2 hours. The tests in this report were all run for 5 minutes.

The tests are not equally relevant - some tests closely simulate how TerraServer works whereas others don't. The following ranks each test's value to TerraServer.

1. The Random Read with 4 overlapped I/Os simulates how the TerraServer web application operates. The TerraServer web application is read-only and primarily performs single row fetches or single key index probes.
2. The Random Write with 4 overlapped I/Os simulates how the TerraServer load processes operate. The TerraServer loaders operate one row at a time within a single transaction. A probe for tile existence, fetch meta data, possibly fetch image data, insert new row, and delete old row is the profile of the TerraServer load programs.
3. The Sequential Read with 4 overlapped I/Os simulates how databases are backed up. The SQL backup utility sequentially read the database being backed up. SQL database restore sequentially reads an on-line backup whenever a server loses a disk and a database is restored. Software mirroring also does sequential reads during the restore of a mirror.
4. Sequential Write with 4 overlapped I/Os simulates how data is restored to a replaced disk. The files are either copied using file system utilities or restored using the SQL Restore utility. Both methods execute large sequential write operations.
5. Random Read with 1 overlapped I/O measures the ability of one drive to perform random read I/O operations without measuring the ability of the controller.
6. Random Write with 1 overlapped I/O measures the ability of one drive to perform random write I/O operations without measuring the ability of the controller.

#### 2.1.2 TerraServer Database Request Playback

The second method was designed to place the Storage Bricks under our real-world workload. We wrote a program, `StressTestFromWebLog.exe` that simulates the database operations performed by the TerraServer web application by reading a TerraServer web log and generating the appropriate requests.

`StressTestFromWebLog` reads the web log files generated on the production web servers. The four production web servers each generate a web log. Each of these logs drives a separate instance of `StressTestFromWebLog` executed on a machine that has the same or more CPU power than the production web servers.

The TerraServer web application runs within the ASP.NET framework and calls SQL stored procedures using ADO.NET. StressTestFromWebLog simulates the ASP.NET environment by using multiple threads to invoke the same SQL stored procedures using ADO.NET as the ASP.NET application does.

The query string information in the web log has enough information to be able to discern the SQL stored procedure parameter values for six of the TerraServer SQL stored procedures that represent 97% of the daily database transactions.

StressTestFromWebLog can be run in one of two modes – *fire hose* or *time synchronized*. In fire-hose mode, the program executes SQL stored procedures as fast as possible. Each instance of the program will run up to 20 requests in parallel. If all 20 SQL processing threads are active, then the program waits for a thread to complete before invoking another parallel thread. This is similar to ASP.NET operation under load.

In time-synchronized mode, StressTestFromWebLog uses the time stamp in the web log to pace the SQL database requests at the same rate they appear in the web log. We monitored the system performance to measure the performance difference between fire hose and time synchronized mode.

## 2.2 Test Results

### 2.2.1 SQLIO Stress Tests

All these tests measure the performance of a single mirrored 250 GB SATA volume. We filled the volume with a 250GB file and ran SQLIO against that file. For random IO this is a worst case scenario since reads have no locality – they are randomized across the entire disk.

The following tables list the **Test** type, disk **Vendor**, **IOps** (I/O Operations per second), **MBps** (Mega-Bytes per second), and **% of Best**.

The column heading identifies what was going on with the system and controllers. **Normal** indicates that the system and controllers were only executing the test. **Ctlr Rebuild** indicates that the RAID controller executing the test was re-building a mirror, but the test was not measuring that mirror. The **Vol Rebuild** means the controller was rebuilding the mirrored volume being measured. The **Vendor** column identifies the drive manufacturer ("WD" for Western Digital.) The **Mirror** column describes the kind of RAID1 volume: (**Software** for Windows 2003 mirroring [MSW2k303], **3ware** for 3Ware controller mirroring [3Ware], and None for no RAID.) We forgot to measure Maxtor Ctlr Rebuild cases so those values are zero.

| Table 1: SQLIO Random 8KB 4-Deep Read Volume. | | | | |
|---|---|---|---|---|
| Vendor | Mirror | Normal IOps | Ctlr Rebuild IOps | Vol Rebuild IOps |
| Maxtor | Software | 146 | 0 | 53 |
|  | 3ware | 137 | 0 | 49 |
|  | None | 72 | 73 | 0 |
| WD | Software | **146** | **145** | 58 |
|  | 3ware | 138 | **144** | 54 |
|  | None | **75** | 74 | 0 |
| % of best | | | | |
| Vendor | Mirror | Normal | Ctlr Rebuild | Vol Rebuild |
| Maxtor | Software | **100%** | 0% | 36% |
|  | 3ware | 94% | 0% | 34% |
|  | None | 96% | 97% | 0% |
| WD | Software | **100%** | **99%** | **40%** |
|  | 3ware | 95% | **99%** | 37% |
|  | None | **100%** | **99%** | 0% |

Table 1 shows the Random Read with 4 overlapped I/Os tests. Windows 2003 mirroring slightly outperforms 3Ware hardware mirroring. Mirror rebuilds reduce Random Read performance by about 70% in both the software and hardware mirror case.

Table 2 shows Random 8KB Write performance with 4 overlapped I/Os. Similar to the read tests, the write tests Windows 2003 software mirrors perform best under normal circumstances. Repairing a broken mirror cuts controller and volume performance almost in ½ for both software and hardware mirroring – even if the volume is not involved in the failure. However, software mirroring outperformed hardware mirroring in most cases.

| Table 2: SQLIO Random 8KB 4-Deep Write Volume. | | | | |
|---|---|---|---|---|
| Vendor | Mirror | Normal IOps | Ctlr Rebuild IOps | Vol Rebuild IOps |
| Maxtor | Software | 117 | 0 | **60** |
|  | 3ware | 117 | 0 | 57 |
|  | None | 116 | 115 | 0 |
| WD | Software | **127** | 79 | **79** |
|  | 3ware | 120 | **83** | 54 |
|  | None | 130 | 132 | 0 |
| % of best | | | | |
| Vendor | Mirror | Normal | Ctlr Rebuild | Vol Rebuild |
| Maxtor | Software | 92% | 0% | **47%** |
|  | 3ware | 92% | 0% | 45% |
|  | None | 89% | 88% | 0% |
| WD | Software | **100%** | 62% | **62%** |
|  | 3ware | 94% | **65%** | 43% |
|  | None | **100%** | **102%** | 0% |

| Table 3: Sequential 64KB 4-Deep Read SATA Volume | | | | |
|---|---|---|---|---|
| Vendor | Mirror | Normal MBps | Ctlr Rebuild MBps | Vol Rebuild MBps |
| Maxtor | Software | 48 | 0 | 6.68 |
|  | 3ware | 33 | 0 | 8 |
|  | None | **48** | **16** | 0 |
| WD | Software | 50 | 0 | **10** |
|  | 3ware | 33 | **14** | 7.5 |
|  | None | **50** | **14** | 0 |
| % of best | | | | |
| Vendor | Mirror | Normal | Ctlr Rebuild | Vol Rebuild |
| Maxtor | Software | 96% | 0% | 13% |
|  | 3ware | 66% | 0% | 16% |
|  | None | 96% | 32% | 0% |
| WD | Software | **100%** | 0% | 20% |
|  | 3ware | 66% | 28% | 15% |
|  | None | **100%** | 28% | 0% |

Table 3 shows the Sequential Read performance (MBps) of 4-deep SQLIO tests. Windows 2003 software mirrors match the performance of no mirroring or failed mirroring. At 50 MBps, it is understandable why "serial" is the first word in the name of the drives.

The Sequential Read test identified a problem with the 3Ware mirroring that seems caps the maximum volume read throughput at 32 MB/sec (525 IOps).

Table 4 shows the bandwidth of the 64KB 4-deep Sequential Write tests. In all cases, the Western Digital drives out performed the Maxtor drives. Windows 2003 software mirroring outperformed 3Ware hardware mirroring except during rebuild where they are equal. Rebuilding a mirror has substantial performance costs. The 3Ware sequential read performance problem disappeared in the write case.

| Table 4: Sequential 64KB 4-Deep Write SATA Volume. | | | | |
|---|---|---|---|---|
| **Vendor** | **Mirror** | **Normal MBps** | **Ctlr Rebuild MBps** | **Vol Rebuild MBps** |
| **Maxtor** | Software | 48 | 0 | 25 |
| | 3ware | 48 | 0 | 10 |
| | None | 48 | 16 | 0 |
| **WD** | Software | 50 | 24 | 24 |
| | 3ware | 50 | 6.6 | 6.6 |
| | None | 50 | 14 | 0 |
| *% of best* | | | | |
| **Vendor** | **Mirror** | **Normal** | **Ctlr Rebuild** | **Vol Rebuild** |
| **Maxtor** | Software | 96% | 0% | 50% |
| | 3ware | 96% | 0% | 20% |
| | None | 96% | 32% | 0% |
| **WD** | Software | 100% | 48% | 48% |
| | 3ware | 100% | 13% | 13% |
| | None | 100% | 28% | 0% |

| Table 5: Random 8KB 1-Deep Read SATA Volume | | | | |
|---|---|---|---|---|
| **Vendor** | **Mirror** | **Normal IOps** | **Ctlr Rebuild IOps** | **Vol Rebuild IOps** |
| **Maxtor** | Software | 76 | 0 | 30 |
| | 3ware | 78 | 0 | 49 |
| | None | 73 | 66 | 0 |
| **WD** | Software | 72 | 49 | 53 |
| | 3ware | 75 | 73 | 53 |
| | None | 74 | 63 | 0 |
| *% of best* | | | | |
| **Vendor** | **Mirror** | **Normal** | **Ctlr Rebuild** | **Vol Rebuild** |
| **Maxtor** | Software | 97% | 0% | 38% |
| | 3ware | 100% | 0% | 63% |
| | None | 99% | 89% | 0% |
| **WD** | Software | 92% | 63% | 68% |
| | 3ware | 96% | 94% | 68% |
| | None | 100% | 28% | 0% |

Table 5 shows 1-deep Random Read performance. Maxtor drives performed about 7% better. Software was slightly slower than hardware mirroring. The test shows the benefit of overlapped I/Os; 4-deep IO has nearly a 100% performance advantage since it can use both disk arms. Like all other tests, mirror rebuild dramatically reduces performance.

Table 6 shows the 1-deep 8KB Random Write performance. Win2003 software mirroring out performed the 3Ware hardware mirroring on Western Digital drives but the performance was the same on Maxtor drives. As in all tests, rebuilding mirrors is an expensive operation.

| Table 6: Random 8KB 1-Deep Write SATA Volume | | | | |
|---|---|---|---|---|
| **Vendor** | **Mirror** | **Normal IOps** | **Ctlr Rebuild IOps** | **Vol Rebuild IOps** |
| **Maxtor** | Software | 116 | 0 | 41 |
| | 3ware | 117 | 0 | 58 |
| | None | 115 | 115 | 0 |
| **WD** | Software | 127 | 0 | 53 |
| | 3ware | 120 | 64 | 52 |
| | None | 130 | 129 | 0 |
| *% of best* | | | | |
| **Vendor** | **Mirror** | **Normal** | **Ctlr Rebuild** | **Vol Rebuild** |
| **Maxtor** | Software | 91% | 0% | 32% |
| | 3ware | 92% | 0% | 46% |
| | None | 88% | 88% | 0% |
| **WD** | Software | 100% | 0% | 42% |
| | 3ware | 94% | 50% | 41% |
| | None | 100% | 99% | 0% |

### 2.2.2 StressTestFromWebLog

We captured the web logs from the four TerraServer production web servers for Monday August 25, 2003. Monday's are traditionally busier days on average than other days of the week. Each log file was copied to a separate Windows 2003 server on the same network switch as the two Storage Brick servers. The Windows 2003 servers physically mimic the production web servers. Each server runs a copy of the StressTestFromWebLog program.

Three of the servers running StressTestFromWebLog and the two Storage Bricks are connected to the same one gigabit network switch. The fourth server (**TS-WEB**) was connected to the building LAN at 100 Mbps and then to the Gbps network switch. As Table 7 shows, this server could not achieve the same throughput as servers directly attached to the gigabit switch.

We executed StressTestFromWebLog in fire-hose mode repeatedly running all four web logs in parallel. The results from each run were very similar with a small standard deviation. Table 7 shows the average database throughput (calls per second) and average database calls handled by each server.

The significant point is that two Storage Bricks are able to sustain average 180 database calls per second per storage server and the pair can execute 360 database calls per second. This is approximately four times the peak weekly load of TerraServer-USA.com. (The server throughput averages 20Mbps limit on the data output rate to the Internet.)

| Table 7: Throughput in fire hose mode. | | |
|---|---|---|
| **Test Server** | **Avg Calls per Sec** | **Avg Db Calls** |
| TS-WEB | 79 | 1,892,377 |
| TWEB1 | 94 | 1,748,869 |
| TWEB3 | 91 | 1,696,663 |
| TWEB4 | 90 | 1,891,869 |
| **Grand Total** | **89** | **1,799,739** |

When the StressTestFromWebLog programs were executed in time-synchronized mode, each simulated web server averaged 22.5 database calls per second. Thus we conclude that two Storage Bricks can handle 4 times the normal TerraServer workload.

Table 8 shows the response time of the five most common database-related web pages (each of these pages invokes a single SQL stored procedure). We measured the time to receive the last byte of data from SQL Server. Table 8 shows for each page, the total number of calls, the average execution time (milliseconds), the maximum observed execution time over all tests (Max ms), and the average maximum time over all tests (average of max ms). The Tile fetch and Image meta-data fetch dominate the traffic. The other three methods – download, famous [places], and imageinfo – are insignificant.

| Table 8: Response time of each web page (and corresponding database stored procedure, measured in milliseconds. | | | | |
|---|---|---|---|---|
| **Web Page** | **Calls** | **Avg ms** | **Max ms** | **Avg Max ms** |
| tile | 1,606,854 | 39 | 16,094 | 7,245 |
| image | 132,900 | 63 | 8,188 | 5,417 |
| download | 1,513 | 160 | 12,484 | 4,481 |
| imageinfo | 199 | 42 | 1,531 | 252 |
| famous | 105 | 9 | 1,781 | 289 |
| **Totals** | **1,741,540** | **41** | **16,094** | **9,581** |

The average duration is 41 ms but there some calls have multi-second delays. We observed that these long delays occurred at the very beginning of each test. We summarize this is caused by SQL Server "waking up," loading its buffer pool, compiling the stored procedures, and handling new connection requests all at once.

## 3 Drive Reliability

We stress tested four Western Digital drives in our office for a month driving them full speed with random 8KB IOs while mounted in an ordinary PC cabinet with no special cooling. This is a more demanding test than placing the drives in a correctly cooled and mechanically stable mounting. The test reported no drive errors.

During the entire experiment we dealt with 32 drives. We saw one drive failure. We ascribe that to the burn-in process – the drive was returned to the vendor for evaluation.

## 4 Comparing Controllers

3ware sells the only controller 66 MHz-64bit PCI SATA controller at present. The 66/64 PCI interface has a theoretical bandwidth 528 MBps (=66x8) whereas the 33/32 PCI busses are limited to 132 MBps (33x4).

The 3ware controllers cost about 70$/port and so add about 25% to the cost of the disk system if you buy large drives, and up to 100% of the cost of the system if you buy small (80GB) drives.

Other companies, notably Highpoint Technologies and Promise offer less expensive 33/32 PCI SATA cards. We evaluated a 345$ 3ware Escalade 8506-4 4-port 66/64 PCI SATA card versus a 71$ Highpoint 1540 33/32 PCI SATA card (branded as a RocketRaid™ card).

Both cards have comparable RAID software and management software on Windows. Both are easy to manage and both have email alerting. 3ware has a background raid scrubber – but otherwise the card's management software seems comparable.

Figure 1 shows the sequential throughput of an 8-disk 3ware Escalade 8508 controller (costing 580$) versus a 4-

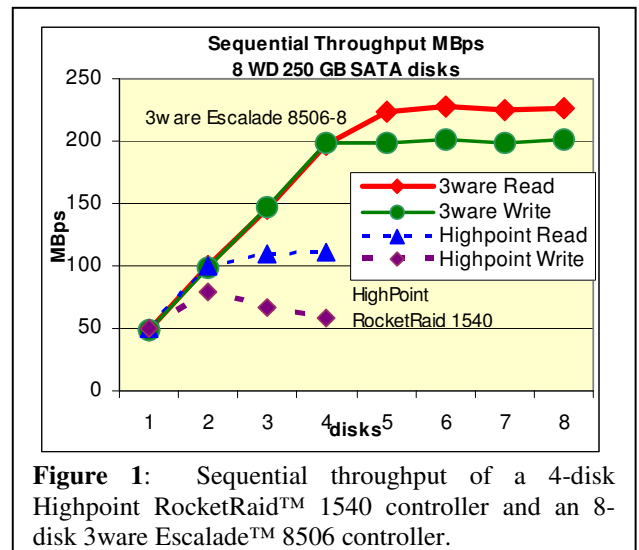

**Figure 1**: Sequential throughput of a 4-disk Highpoint RocketRaid™ 1540 controller and an 8-disk 3ware Escalade™ 8506 controller.

disk Highpoint 1540 RocketRaid controller (costing 70$). All tests are 4-deep. Highpoint Sequential reads saturate at 110 MBps and 2 disks. Sequential writes saturate at 79MBps and 2 disks. 3ware saturates at 225 MBps read and 200MBps write. The combination of a faster PCI bus (66 MHz and 64 bits wide) and more powerful controller gives the 3ware 2 to 3 times the sequential performance of the 7x less expensive Highpoint card.

The situation on random IO is somewhat different. Figure 2 shows that the Highpoint RocketRaid delivers better performance than the 3ware card on up to 4 disks. Random IO speed is not limited by the PCI Bus speed.

The appendix has the detailed measurements but Figure 2 summarizes the comparison.

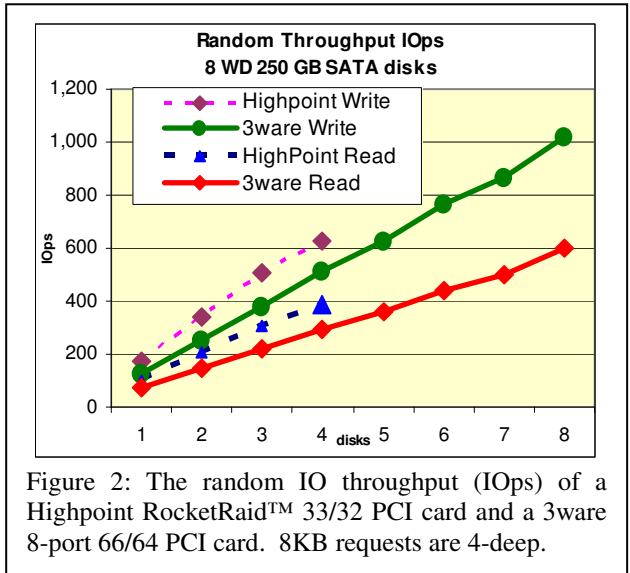

Figure 2: The random IO throughput (IOps) of a Highpoint RocketRaid™ 33/32 PCI card and a 3ware 8-port 66/64 PCI card. 8KB requests are 4-deep.

In short, the Highpoint card delivers superior random IO performance but saturates at 100MBps on sequential IO (the bandwidth of 2 disks) while the 3ware card delivers approximately twice the sequential IO performance and saturates at 200 MBps (the bandwidth of 4 disks).

Our conclusion from this is: if you have a random IO workload, the inexpensive SATA controllers are adequate and can save about 20% to 100% per drive when compared to the 3ware controllers. If you have a heavy sequential workload, the 3ware card has about 100% performance advantage – but 4 disks will saturate the card, so you should buy multiple 4-port cards rather than putting 8 or 12 disks on one card.

## 5 Conclusion

The StressTestFromWebLog testing verified that our Storage Bricks with SATA drives can easily accommodate the TerraServer web application work load regardless of which SATA drive we select, which RAID controller we select, or how RAID1 mirroring is implemented. These choices can be made for economic, ease-of-management, and availability characteristics without regard to performance.

From these experiments we conclude:

- **Both the Western Digital and Maxtor drives were reliable**.

- **The Western Digital drives performed better on most tests** including the random I/O tests which are the most similar to the TerraServer application workload and had comparable prices. Therefore, we selected Western Digital 250GB drives for the Storage Bricks.

- **3ware controllers are superior for high-speed sequential.**

- **For random IO the less expensive Highpoint controllers have superior performance,** and they are 7x less expensive per port. So if there are enough PCI slots, we recommend using the Highpoint cards for workloads that do not need high-speed sequential IO.

- **We found no performance difference between a 2x4-port RAID controller and a 1x8-port RAID controller?** We found no availability or management difference between the 4-port and 8-port 3Ware RAID controller. Using the 4-port controller unnecessarily consumes 64-bit PCI slots on the motherboard and is more expensive per port than the 8-port controller. Therefore, we selected the 8-port 3Ware SATA RAID controller for all Storage Bricks.

- **There are performance and manageability differences between hardware RAID1 and software RAID1, but we are not convinced which is better for TerraServer in the long run.**
The two management interfaces and availability profiles are very different. We did not completely explore them during our tests. 3Ware offers a web based management and monitoring environment that offers a number of interesting features – automatic e-mail and pager alarms, background data integrity checking – that are not available "out of the box" with Windows 2003 software mirroring. However, Windows 2003 is integrated into the Windows Computer Management application, emits serious events to the Event Log, and can be monitored by the Microsoft Operations Management (MOM) environment.

Software mirroring allows us to mirror and stripe data across 3Ware controllers whereas a 3Ware RAID-1 mirror is implemented by a single controller. This gives the availability edge to software mirroring. Although in practice, we have never seen a 3Ware controller fail. Changing a mirror configuration with hardware RAID requires an OS reboot (for both Highpoint and 3ware) and 3ware configuration is done in a "DOS" phase during reboot. Software mirroring allows online change to the configuration.

We decided that more study is required. We intend to run half of our Storage Bricks using software mirroring and half with hardware mirroring to see if any long-term benefits or issues with the alternative technologies.

- **Two Storage Bricks can service 400% of the normal peak weekly TerraServer application load.** We executed 97% of the actual database requests captured on August 25[th,] 2003. The two Storage Bricks could process the 24 hours of captured database transactions within less than 7 hours. Even with one simulated web server handicapped by its network connection, each simulated web server executed an average of 89 database calls per second when run in fire hose mode. When run in time synchronization mode, the database calls per second averaged 22.5 per second. Thus we believe that two storage bricks can handle 4 times the normal Monday workload.

Based on these tests, we have decided to purchase six Storage Bricks with the following configuration:
- 3u rack mountable chassis
- Dual Xeon 2.4ghz hyper threaded processors
- SuperMicro motherboard
- 4 GB of RAM
- 2 3Ware 8-port 8506-8 SATA RAID controllers
- 16 Western Digital 250gb SATA drives
- 2 on-board 10/100/1000 Ethernet NICs

Three Storage Bricks will be configured with Windows 2003 RAID-1 mirroring. The other three Storage Bricks will be configured using 3Ware hardware mirroring. While we will devise tests to measure our satisfaction with the manageability and availability with software mirroring versus hardware mirroring, we need to get on with the project. Therefore, we believe that part of the TerraServer Storage Brick project will be to evaluate and compare software mirroring to hardware mirroring in a production application.

# 7 Appendix:

The detailed measurements from the comparison of the 3Ware and Highpoint SATA controllers on 1 to 4 disks (the 3ware controller can manage up to 8 disks and has a 4x faster PCI interface.. Tests done using SQL IO, 4-deep, 256KB sequential 8KB random IOs.

| | | | | Vendor | |
|---|---|---|---|---|---|
| rand/seq | r/w | disks | Rate | 3ware | Highpoint |
| Random | Read | 1 | MBps | 0.6 | 0.8 |
| | | | IOps | 73.8 | 102.8 |
| | | 2 | MBps | 1.2 | 1.6 |
| | | | IOps | 148.2 | 205.4 |
| | | 3 | MBps | 1.7 | 2.4 |
| | | | IOps | 220.0 | 305.6 |
| | | 4 | MBps | 2.3 | 3.0 |
| | | | IOps | 292.2 | 386.7 |
| | | 5 | MBps | 2.8 | |
| | | | IOps | 362.0 | |
| | | 6 | MBps | 3.4 | |
| | | | IOps | 440.8 | |
| | | 7 | MBps | 3.9 | |
| | | | IOps | 496.7 | |
| | | 8 | MBps | 4.7 | |
| | | | IOps | 600.5 | |
| | Write | 1 | MBps | 1.0 | 1.4 |
| | | | IOps | 126.1 | 176.5 |
| | | 2 | MBps | 2.0 | 2.6 |
| | | | IOps | 256.4 | 336.7 |
| | | 3 | MBps | 3.0 | 4.0 |
| | | | IOps | 382.1 | 509.1 |
| | | 4 | MBps | 4.0 | 4.9 |
| | | | IOps | 514.8 | 625.4 |
| | | 5 | MBps | 4.9 | |
| | | | IOps | 626.1 | |
| | | 6 | MBps | 6.0 | |
| | | | IOps | 763.7 | |
| | | 7 | MBps | 6.8 | |
| | | | IOps | 867.3 | |
| | | 8 | MBps | 8.0 | |
| | | | IOps | 1,018.3 | |
| Sequential | Read | 1 | MBps | 49.1 | 49.7 |
| | | | IOps | 196.3 | 795.9 |
| | | 2 | MBps | 99.7 | 99.4 |
| | | | IOps | 398.8 | 1,590.6 |
| | | 3 | MBps | 146.3 | 109.2 |
| | | | IOps | 585.4 | 1,746.8 |
| | | 4 | MBps | 197.5 | 111.8 |
| | | | IOps | 790.0 | 1,788.6 |
| | | 5 | MBps | 223.9 | |
| | | | IOps | 895.7 | |
| | | 6 | MBps | 227.2 | |
| | | | IOps | 908.9 | |
| | | 7 | MBps | 224.9 | |
| | | | IOps | 899.5 | |
| | | 8 | MBps | 226.4 | |
| | | | IOps | 905.6 | |
| | Write | 1 | MBps | 49.1 | 49.6 |
| | | | IOps | 196.2 | 794.1 |
| | | 2 | MBps | 99.3 | 78.7 |
| | | | IOps | 397.2 | 1,259.9 |
| | | 3 | MBps | 147.7 | 66.9 |
| | | | IOps | 591.0 | 1,070.6 |
| | | 4 | MBps | 198.5 | 58.6 |
| | | | IOps | 793.8 | 937.3 |
| | | 5 | MBps | 198.5 | |
| | | | IOps | 794.1 | |
| | | 6 | MBps | 201.4 | |
| | | | IOps | 805.5 | |
| | | 7 | MBps | 198.7 | |
| | | | IOps | 794.7 | |
| | | 8 | MBps | 201.2 | |
| | | | IOps | 804.8 | |